\documentstyle[aps,prd,epsfig]{revtex}

\def \be{\begin{equation}}
\def \ee{\end{equation}}
\def \bea{\begin{eqnarray}}
\def \eea{\end{eqnarray}}

\def \s2{\sqrt 2}

\begin{document}

\draft

\title{Relativistic hydrodynamics on spacelike and null surfaces:
Formalism and computations of spherically symmetric spacetimes}

\author{Philippos Papadopoulos and Jos\'e A. Font}

\address{Max-Planck-Institut f\"ur Gravitationsphysik \\
Albert-Einstein-Institut \\
Schlaatzweg 1, D-14473, Potsdam, Germany
}

\date{\today}

\maketitle

\begin{abstract}

We introduce a formulation of Eulerian general relativistic
hydrodynamics which is applicable for (perfect) fluid data prescribed
on either spacelike or null hypersurfaces. Simple explicit expressions
for the characteristic speeds and fields are derived in the general
case. A complete implementation of the formalism is developed in the
case of spherical symmetry. The algorithm is tested in a number of
different situations, predisposing for a range of possible
applications.  We consider the Riemann problem for a polytropic gas,
with initial data given on a retarded/advanced time slice of Minkowski
spacetime. We compute perfect fluid accretion onto a Schwarzschild
black hole spacetime using ingoing null Eddington-Finkelstein
coordinates.  Tests of fluid evolution on dynamic background include
constant density and TOV stars sliced along the radial null
cones. Finally, we consider the accretion of self-gravitating matter
onto a central black hole and the ensuing increase in the mass of the
black hole horizon.

\end{abstract}


\section{Introduction}
\label{sec:intro}

The theoretical understanding of the detailed dynamics of black hole
interactions with matter is today a key research activity, both in the
gravitational wave~\cite{thorne} and high-energy
astrophysics~\cite{rees} communities. Such research targets the
interpretation of data obtained (or anticipated) in a number of
distinct observational windows, from high resolution spectra of
regions suspected of harboring black holes, e.g., with satellite
experiments like the Rossi X-ray Timing Explorer (RXTE)~\cite{RXTE},
to broad band gravitational wave detection efforts, e.g., the Laser
Interferometric Gravitational Observatory (LIGO)~\cite{LIGO}.
Computations within the Newtonian paradigm have reached high levels of
sophistication (see, e.g.,~\cite{ruffert}), offering important clues
and support for further development of astrophysical models.  Clearly,
computations within the framework of general relativity would be
highly desirable. Several efforts are underway for meeting the
difficult challenge of solving the Einstein equations in full
generality (e.g.,~\cite{inrg0,inrg1,inrg2,gc,centrella}).

Yet the task before the computational scientists is daunting, as the
underlying theory is rich in conceptual and technical
complications. This attribute of relativistic gravity, led in the
pre-60's period to a certain confusion with respect to the physical
content of the theory, e.g., as to whether gravitational waves exist
at all. The conundrum was finally resolved with the introduction and
masterly manipulation of the characteristic initial value problem
(CIVP) by Bondi and Sachs~\cite{bondi-paper,sachs-paper}. Important
features of the CIVP suggest it may be serve as a valuable tool in the
effort of studying interesting black hole spacetimes (a review of the
development of algorithms based on the CIVP is given
in~\cite{LIVING}). We highlight, in particular, the long term
stability of non-linear numerical evolution schemes based on null
coordinates, shown for regular spacetimes in
axisymmetry~\cite{bondi-code}, and black hole spacetimes in full
3D~\cite{3d-code}. Additionally, the schemes are of low computational
cost per spacetime point, an issue which acquires significance when
stability problems are controlled. Both features ultimately derive
from the gauge properties of the null foliation, which captures
directly the wave degrees of freedom as a propagation equation for a
complex function, with all other relevant equations converted into
ODE's to be integrated along the characteristic surface. The CIVP
formulation of the Einstein equations hence offers, with the inclusion
of appropriate matter dynamics (e.g., in the form of perfect fluid
hydrodynamics), among others, the possibility of accurate studies of
generic black-hole-matter interactions.

The aim of this paper is to present and test a general formulation of
the general relativistic hydrodynamic (GRH) equations for ideal
fluids, appropriate for numerical work, which is suited, but not
restricted, to integration on spacetimes foliated with null
hypersurfaces. The {\em form invariance} of the approach with respect
to the nature of the foliation implies that existing work on highly
specialized techniques for fluid dynamics can be adopted with minimal
modifications. In our program of studying black hole matter
interactions we have already used such techniques in estimating
gravitational radiation from bulk fluid accretion~\cite{bp-hydro} in
the black hole perturbation limit. The developments here constitute a
first step in extending that program into the fully non-linear regime.

The paper contains four main sections.  Section~\ref{sec:formalism}
presents the essential formal elements of a new prescription for
solving GRH, i.e., a choice of conserved and primitive variables and
their relationship, along with a diagonalization of the Jacobian
matrix and explicit expressions for its eigenvalues and
eigenvectors. In Section~\ref{sec:general-1d} we formulate the problem
of coupled evolution of matter and geometry fields in spherical
symmetry. The analog of the Tolman-Oppenheimer-Volkoff (TOV) equation
is constructed and integrated along the null cone. Those solutions are
used for consistency and convergence tests. Those are demonstrated,
along with some relevant details of the numerical implementation, in
Section~\ref{sec:alg-tests}. That Section also discusses the important
issue of shock-capturing, explored for fluid data posed on a null
surface (in flat spacetime). In Section~\ref{sec:accretion} we turn to
black hole spacetimes and spherical accretion. The test fluid limit is
used as a test-bed, and we proceed to a preliminary study of self
gravitating accretion.

We use geometrized units ($G=c=1$) and the metric conventions 
of~\cite{MTW}. Boldface letters (and capital indices) denote 
vectors (and their components) in the fluid state space.

\section{Formalism for general relativistic hydrodynamics}
\label{sec:formalism}

\subsection{Prelude to the formal developments}

The traditional approach for relativistic hydrodynamics on spacelike
hypersurfaces is based on Wilson's pioneering work~\cite{wilson72}.
In this formalism the equations were originally written as a set of
advection equations. This approach sidesteps an important guideline
for the formulation of non-linear hyperbolic systems of equations,
namely the preservation of their {\it conservation form}. This feature
is necessary to guarantee correct evolution in regions of sharp
entropy generation. Nevertheless, in the absence of such features,
non-conservative formulations are equivalent to conservative ones. The
approach is then simpler to implement and has been widely used over
the years in a number of astrophysical scenarios. On the other hand,
a numerical scheme in conservation form allows for shock-capturing,
i.e., it guarantees the correct Rankine-Hugoniot (jump) conditions
across discontinuities. Writing the relativistic hydrodynamic
equations as a system of conservation laws, identifying the suitable
vector of unknowns and building up an approximate Riemann solver has
permitted the extension of modern {\it high-resolution
shock-capturing} (HRSC in the following) schemes from classical fluid
dynamics into the realm of relativity~\cite{mim}. The main theoretical
ingredients to construct such a scheme in full general relativity can
be found in~\cite{betal97}. An up-to-date collection of different
applications of HRSC schemes in relativistic hydrodynamics is
presented in~\cite{jcam}.  We will return to these schemes in
section~\ref{sec:alg-tests} below.

Non finite-difference methods have also been applied recently to
compute relativistic flows, most notably, {\it pseudo-spectral
methods} and {\it smoothed particle hydrodynamics} (SPH) techniques.
Although pseudo-spectral methods have enhanced accuracy in smooth
regions of the solution, the correct modeling of discontinuous
solutions is still their main drawback~\cite{Eric1}.  Recently,
however, progress has been achieved in this direction by using
multi-domain decomposition techniques~\cite{meudon1}.  On the other
hand, SPH, as any other particle method, suffers from being
dissipative when resolving steep gradients~\cite{Mann}. In spite of
this, it has recently proven to work in the ultra-relativistic
regime~\cite{cm97}. Additionally, it has been shown that it is
possible to generalize SPH methods to hyperbolic systems other than
the Euler equations~\cite{spi}

Procedures for integrating various forms of the hydrodynamic equations
on {\em null hypersurfaces} have been presented
before~\cite{isaacson83} (see~\cite{bishop99} for a recent
implementation). This approach is geared towards smooth isentropic
flows. A Lagrangian method, applicable in spherical symmetry, has been
presented by~\cite{jmiller}. Recent work in~\cite{ddv} includes an
Eulerian, non-conservative, formulation for general fluids in null
hypersurfaces and spherical symmetry, including their {\em matching}
to a spacelike section. Here we show that the GRH equations can be
formulated in a conservative and {\em form invariant} way (i.e.,
irrespective of the spacelike or null nature of the foliation) for an
arbitrary three dimensional spacetime and any perfect fluid with
polytropic equation of state.

\subsection{The formulation}

\subsubsection{Variables and evolution equations}

We consider the relativistic conservation equations (continuity
equation and Bianchi identities) upon introducing an explicit
coordinate chart $(x^{0},x^{i})$, i.e., 
\begin{eqnarray}
\label{initial1}
\frac{\partial}{\partial x^{\mu}} \sqrt{-g} J^{\mu} & = & 0 \, , \\
\frac{\partial}{\partial x^{\mu}} \sqrt{-g} T^{\mu\nu} & = & - \sqrt{-g}
\Gamma^{\nu}_{\mu\lambda} T^{\mu\lambda} \, ,
\label{initial2}
\end{eqnarray}
where the scalar $x^{0}$ represents a foliation of the spacetime with
hypersurfaces (coordinatized by $(x^{1},x^{2},x^{3})$), with the matter
current and stress energy tensor for a perfect fluid given by $J^{\mu}
= \rho u^{\mu} $, $ T^{\mu\nu} = \rho h u^{\mu} u^{\nu} + p
g^{\mu\nu}$, where $p$ is the pressure, $u^{\mu}$ is the fluid four
velocity and $h = 1 + \varepsilon + p/\rho$ is the relativistic
specific enthalpy. 

We do not restrict the foliation to be spacelike (that is, the level
surfaces of $x^{0}$ may be also null). We define the coordinate
components of the four-velocity $u^{\mu} = (u^{0}, u^{i})$. The
velocity components $u^{i}$, together with the rest-frame density and
internal energy, $\rho$ and $\varepsilon$, provide a unique
description of the state of the fluid and will be called (following
common usage) the {\em primitive} variables. They constitute a vector
in a five dimensional space $ {\bf w}^{A} = (\rho, u^{i},
\varepsilon)$. The index $A$ is taken to run from zero to four,
coinciding for the values (1,2,3) with the coordinate index $i$. We
define the initial value problem for
equations~(\ref{initial1},\ref{initial2}) in terms of another vector
in the same fluid state space, namely the {\em conserved variables},
${\bf U}^{B}$, individually denoted $(D,S^{i},E)$~\cite{projection},
\begin{eqnarray}
\label{eq:D}
D     & = & {\bf U}^{0} = J^{0}  = \rho u^{0} \, ,  \\ 
S^{i} & = & {\bf U}^{i} = T^{0i} = \rho h u^{0} u^{i} + p g^{0i} \, ,  \\
E     & = & {\bf U}^{4} = T^{00} = \rho h u^{0} u^{0} + p g^{00} \, . 
\label{eq:E}
\end{eqnarray}
With those definitions the equations will take the standard
conservation law form\cite{non-conservation},
\begin{equation}
\partial_{x^{0}} (\sqrt{-g} {\bf U}^{A}) 
+ \partial_{x^{j}} (\sqrt{-g} {\bf F}^{jA})
= {\bf S}^{A} \, ,
\label{eq:cons-law}
\end{equation}
where $\sqrt{-g}$ is the volume element associated with the
four-metric and we defined the flux vectors ${\bf F}^{jA}$ and the
source terms ${\bf S}^{A}$ (which depend only on the metric, its
derivatives and the undifferentiated stress energy tensor),
\begin{eqnarray}
{\bf F}^{j0} & = &  J^{j}  = \rho u^{j} \, , \nonumber \\ 
{\bf F}^{ji} & = &  T^{ji} = \rho h u^{i} u^{j} + p g^{ij} \, , \nonumber \\
{\bf F}^{j4} & = &  T^{j0} = \rho h u^{0} u^{j} + p g^{0j} \, ,
\label{eq:fluxes} \\ \nonumber
\\ 
{\bf S}^{0} & = & 0 \, , \nonumber \\
{\bf S}^{i} & = & - \sqrt{-g} \, \Gamma^{i}_{\mu\lambda} T^{\mu\lambda}  
\, , \nonumber \\
\label{eq:sources}
{\bf S}^{4} & = & - \sqrt{-g} \, \Gamma^{0}_{\mu\lambda} T^{\mu\lambda} \, .
\end{eqnarray}

The state of the fluid is uniquely described using either vector of
variables, i.e., ${\bf U}^{A}$ or ${\bf w}^{A}$, and one can be
obtained from the other, as will be shown later, via the
definitions~(\ref{eq:D})-(\ref{eq:E}) and the use of the normalization
condition for the velocity vector,
\begin{equation}
g_{\mu\nu} u^{\mu} u^{\nu} = -1 \, .
\end{equation} 

Specification of ${\bf U}^{A}$ on the initial hypersurface, together
with an equation of state (EOS) $p = p(\rho,\varepsilon)$, followed by
a recovery of the primitive variables leads to the computation of the
fluxes and source terms. Hence, the first time derivative of the data
is obtained, which then leads to the formal propagation of the
solution forward in time. No continuity of the data is required, since
in practice the evolution is achieved with the (possibly approximate)
solution of local Riemann problems.

\subsubsection{Local characteristic structure of the equations}

Utilizing the machinery of modern hydrodynamical methods, i.e., HRSC
schemes, to integrate the previous equations requires the
investigation of their local characteristic structure. We proceed here
with this analysis. For this purpose we temporarily ignore the
inhomogeneous part of the system.

Introducing the Jacobian matrices
\begin{equation}
{\bf B}^{jA}_{B} =  \frac{\partial{\bf F}^{jA} }{\partial {\bf U}^{B}} \, ,
\end{equation}
the system can be written in quasi-linear form as
\begin{equation}
\partial_{x^{0}} {\bf U}^{A} 
+ {\bf B}^{jA}_{B}  \partial_{x^{j}} {\bf U}^{B} = 0 \, .
\label{B-system}
\end{equation}

The characteristic speeds (eigenvalues) and eigenvectors of the
system, in the $j$-th direction, are then given by the solution of the
algebraic problems (we now omit the vector indices of the fluid
space)
\begin{eqnarray} 
& & det({\bf B}^{j} - \lambda^{j} {\bf I}) =  0 \, ,  \\
& & ({\bf B}^{j} - \lambda^{j} {\bf I}) {\bf r}^{j} = 0 \, ,
\end{eqnarray}
where $\lambda^{j}$ denotes the characteristic speed in the
direction $j$, ${\bf r}^{j}$ the corresponding eigenvector and
${\bf I}$ denotes the unit matrix.

The strong coupling of the elements of the ${\bf B}$ matrix via the
normalization condition hinders the analysis of the eigenvalue problem
in terms of the conserved variables ${\bf U}$.  The customary method
to proceed (see~\cite{font1}), is to analyze the problem in terms of
the primitive variables. Indeed, the suitable choice of the primitive
variables, such that the eigenvalue problem may be solved explicitly,
is of great practical value.

With a choice of primitive variables ${\bf w}$, the quasi-linear
system~(\ref{B-system}) is rewritten as 
\begin{equation}
{\bf A}^{0} \partial_{x^{0}} {\bf w}
+  {\bf A}^{j} \partial_{x^{j}} {\bf w} = 0 \, ,
\label{A-system}
\end{equation}
where 
\begin{eqnarray}
{\bf A}^{0} & = & \frac{\partial{\bf U}}{\partial {\bf w}} \, , \\
{\bf A}^{j} & = & \frac{\partial{\bf F}^{j} }{\partial {\bf w}} \, . 
\end{eqnarray}

Hence, upon analyzing the algebraic system 
\begin{eqnarray} 
& & det({\bf A}^{j} - \bar{\lambda}^{j} {\bf A}^{0}) =  0 \, , \\
& & ({\bf A}^{j} - \bar{\lambda}^{j} {\bf A}^{0}) \bar{{\bf r}}^{j} = 0 \, ,
\end{eqnarray}
elementary algebra establishes that $\bar{\lambda}^{j} = \lambda^{j}$
and the eigenvectors of matrix ${\bf B}^{j}$ are given by $\bar{{\bf
r}}^{j} = {\bf A}^{0} {\bf r}^{j}$.

We now proceed to diagonalize our system, with the choice of primitive
variables ${\bf w}^{B} = (\rho,u^{i},\varepsilon)$, and specializing
the derivation to the case of a perfect fluid EOS, $p=(\Gamma-1)
\rho\varepsilon$, with $\Gamma$ being the (constant) adiabatic index
of the fluid. The procedure can be generalized to an arbitrary EOS and
details will be given elsewhere.

The matrices ${\bf A}$ are in this case
\begin{center}
\[ {\bf A}^{0} = \left[ \begin{array}{ccc}
u^0 & \rho \mu_i & 0  \\
hu^0u^k + (\Gamma-1)\varepsilon g^{0k} & 
\rho h (u^0\delta^k_i+u^k\mu_i) & 
\rho\Gamma u^0u^k+(\Gamma-1)\rho g^{0k} \\
h(u^0)^2+(\Gamma-1)\varepsilon g^{00} &
2\rho h u^0\mu_i & 
\rho\Gamma (u^0)^2 +(\Gamma-1)\rho g^{00} 
\\
\end{array} \right] \, , \] 
\end{center}
\begin{center}
\[ {\bf A}^{i} = \left[ \begin{array}{ccc}
u^i & \rho \delta^i_j & 0 \\
hu^ku^i + (\Gamma-1)\varepsilon g^{ki} &
\rho h \delta^k_j u^i + \rho h u^k \delta^i_j &
\rho\Gamma u^ku^i+(\Gamma-1)\rho g^{ki} \\
hu^0u^i+(\Gamma-1)\varepsilon g^{0i} &
\rho h u^0\delta^i_j + \rho hu^i\mu_j &
\rho\Gamma u^0u^i +(\Gamma-1)\rho g^{0i} 
\\
\end{array} \right] \, , \] 
\end{center}
\noindent
with $\mu_i\equiv\frac{\partial u^0}{\partial u^i}=-\frac{u_i}{u_0}$.
For a given coordinate direction, which we label `1' (e.g., the $x^{1}$
direction), the matrix ${\bf A}^1- \lambda^{1} {\bf A}^0$ reads
\begin{center}
\[ {\bf A}^{1} -\lambda^{1} {\bf A}^{0} = \left[ \begin{array}{ccc}
a & \rho b_i & 0 \\
hu^ka + (\Gamma-1)\varepsilon c^k &
\rho h \delta^k_i a + \rho h u^k b_i &
\rho\Gamma u^ka+(\Gamma-1)\rho c^k \\
hu^0a+(\Gamma-1)\varepsilon d &
\rho h u^0b_i + \rho ha\mu_i &
\rho\Gamma u^0a +(\Gamma-1)\rho d
\\
\end{array} \right] \, , \]  
\end{center}
\noindent
where the following shorthand notation is used
\begin{eqnarray}
a   &\equiv& u^1-\lambda^{1} u^0, 
\hspace{1.15cm}
b_i \equiv \delta^1_i - \lambda^{1} \mu_i \\
c^k &\equiv& g^{k1} - \lambda^{1} g^{0k},
\hspace{1cm} 
d   \equiv g^{01} - \lambda^{1} g^{00}.
\end{eqnarray}
\noindent
The {\it eigenvalues} of that matrix are
\begin{equation}
\lambda^{1}_0 = \frac{u^1}{u^0} \mbox{\,\,\,\,(triple)} \, ,
\label{lambda0}
\end{equation}  
\noindent
and
\begin{eqnarray}                                            
\lambda^{1}_{\pm} &=& \frac{1}{1-c_{s}^{2}(1-{\cal L})}
\left[ {\cal M}c_{s}^2 + v^1(1-c_{s}^{2}) \mp 
c_s \sqrt{c_s^2{\cal M}^2 + v^1(1-c_s^2)(2{\cal M}-{\cal L}v^1) +
{\cal N}(1-c_s^2(1-{\cal L}))}
\right] \, ,
\label{lambdapm}
\end{eqnarray}
\noindent
where all the metric information is encoded in the expressions
${\cal L}, {\cal M}$ and ${\cal N}$, 
\begin{eqnarray}
{\cal L}  & \equiv& -\frac{g^{00}}{(u^0)^2} , \,
\hspace{1cm}
{\cal M}  \equiv -\frac{g^{01}}{(u^0)^2} , \,
\hspace{1cm}
{\cal N}  \equiv  \frac{g^{11}}{(u^0)^2}. 
\end{eqnarray}
\noindent
Additionally, $v^1 \equiv \frac{u^1}{u^0}$
and $c_s$ is the local sound speed satisfying
\begin{eqnarray}
h c^2_s = \chi +\frac{p}{\rho^2}\kappa \, ,
\end{eqnarray}
\noindent
with $\chi=\frac{\partial p}{\partial\rho}=(\Gamma-1)\varepsilon$ and
$\kappa=\frac{\partial p}{\partial\varepsilon}=(\Gamma-1)\rho$. A
complete set of {\it right-eigenvectors} is given by
\begin{eqnarray}
{\bf r}_{0,1} = u^0(1,u^1,u^2,u^3,u^0) \, ,
\end{eqnarray}
\begin{eqnarray}
{\bf r}_{0,2} = (u^0+\rho\mu_{32}, 
u^0u^1+\rho h u^1\mu_{32},
u^0u^2+\rho h(u^2\mu_{32} - u^0 b_3), 
u^0u^3+\rho h(u^3\mu_{32} + u^0 b_2),
(u^0)^2+2\rho hu^0\mu_{32}) \, ,
\end{eqnarray}
\begin{eqnarray}
{\bf r}_{0,3} = (u^0+\rho\mu_{23}, 
u^0u^1+\rho h u^1\mu_{23},
u^0u^2+\rho h(u^2\mu_{23} + u^0 b_3), 
u^0u^3+\rho h(u^3\mu_{23} - u^0 b_2),
(u^0)^2+2\rho hu^0\mu_{23}) \, ,
\end{eqnarray}
\[
{\bf r}_{\pm} =
\left[ \begin{array}{c}
-\frac{\rho}{\varepsilon}u^0B + \frac{\rho\Gamma}{h}K
\\ \\
-\frac{\rho}{\varepsilon}u^0Bu^1h + \rho\Gamma[A(u^0c^1-ag^{01})+u^1K]
\\ \\
-\frac{\rho}{\varepsilon}u^0Bu^2h + \rho\Gamma[A(u^0c^2-ag^{02})+u^2K]
\\ \\
-\frac{\rho}{\varepsilon}u^0Bu^3h + \rho\Gamma[A(u^0c^3-ag^{03})+u^3K]
\\ \\ 
-\frac{\rho}{\varepsilon}(u^0)^2{\tilde B} + \rho\Gamma[2u^0K-aAg^{00}]
\\ \end{array}
\right] \, ,
\]
\noindent
with the definitions
\begin{eqnarray}
A          &\equiv& \frac{u^0-u^i\mu_i}{d-c^i\mu_i}, 
\hspace{1.1cm}
{\tilde A} \equiv \frac{\Gamma aA}{\Gamma-1}, \\
B          &\equiv& 1 + {\tilde A},
\hspace{1.7cm}
{\tilde B} \equiv 1 +h{\tilde A}, \\
K          &\equiv& \mu_i(Ac^i-u^i),
\hspace{0.5cm}
\mu_{ij}   \equiv \mu_i b_j -\mu_j b_i.
\end{eqnarray}
\noindent
Note that the $\pm$ dichotomy in the last two eigenvectors is implicit
in the corresponding non-degenerate ($\pm$) eigenvalues through the
variables $a$, $c^i$ and $d$.

The spectral decomposition given above applies to a chosen direction
$j$. Since $j$ is arbitrary, to obtain similar expressions for the
remaining directions, it suffices to specialize them accordingly,
e.g., obtain the eigenvalues from expressions~(\ref{lambda0}) and
(\ref{lambdapm}) with substitution of the desired direction, and
permutation of the corresponding eigenvectors.

\subsubsection{Relations between variables}
\label{primitives}

So far the developments have been completely general. We specialize
here the discussion to null coordinate systems. 

\begin{figure}[t]
\centerline{\epsfig{file=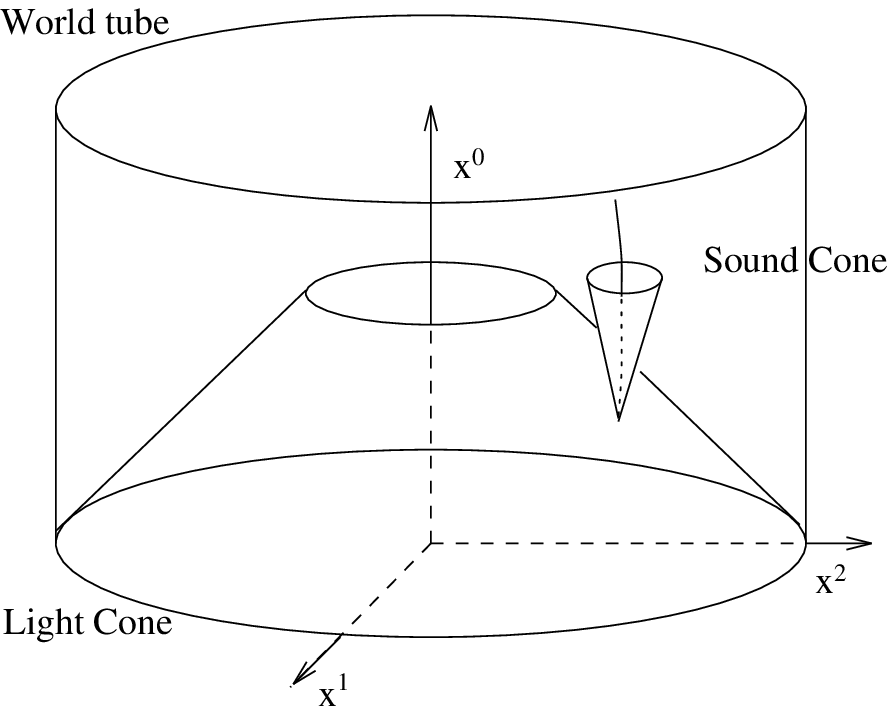,width=2.5in,height=2.0in}}
\vspace{0.2in}
\caption{Fluid equations of state are typically obeying causality,
i.e., for equilibrium configurations the sound cone is contained
within the light cone. In this case specifying fluid data on a null
surface (here depicted as an advanced time null cone centered on the
origin of coordinates) constitutes a Cauchy problem for the fluid. An
example sound cone is depicted as a narrow forward cone, along with a
worldline of a fluid element (in this case sub-sonic flow). This state
of affairs would persist for an arbitrary curved spacetime and any
null surface.}
\label{fig:cones}
\end{figure}

For the EOS commonly accepted, the propagation speeds of fluid signals
are always sub-luminal. In addition, the bulk flow is always assumed
to be a timelike vector field. Hence, the Cauchy initial value problem
for the fluid is well defined for data given on a null hypersurface
(see Fig.~\ref{fig:cones}).

While the numerical algorithm updates the vector of conserved
quantities ($D,S^i,E$), we make extensive use of the {\it primitive}
variables ($\rho,u^i,\varepsilon$). Those would appear repeatedly in
the solution procedure: in the characteristic fields, in the solution
of the Riemann problem and in the computation of the numerical fluxes
(see below).  Hence, it is necessary to specify a procedure for
recovering them from the conserved quantities. In the spacelike case
the relation between the two sets of variables is implicit. An example
of an iterative algorithm to recover the primitive variables in this
situation can be found in~\cite{mm2}.  In the null case, the procedure
of connecting primitive and conserved variables turns out to be {\em
explicit} for a polytropic EOS. This is a direct consequence of the
condition $g^{00}=0$ which characterizes null
foliations~\cite{sachs-paper} and leads to algebraic simplifications
in the normalization expression $g^{\mu\nu}u_{\mu}u_{\nu}=-1$.

Assuming, hence, a perfect fluid EOS, the internal energy,
$\varepsilon$, can be directly obtained in terms of the conserved
quantities as the positive solution to a binomial equation, more
precisely
\begin{equation}
\varepsilon = \frac{\Lambda^2}{D^2+D\sqrt{D^2+\Gamma(2-\Gamma)\Lambda^2}} \, ,
\end{equation}
where
\begin{equation}
\Lambda^2=-D^2-g_{00}E^2-2g_{0i}S^iE-g_{ij}S^iS^j \, .
\end{equation}
Once $\varepsilon$ is known the rest of the {\it primitive} variables
follow, e.g., $ h = 1 + \Gamma \varepsilon$, 
$\rho = D^2 h / E$, and
\begin{equation}
u^i  =  \frac{S^i-pg^{0i}}{D(1+\Gamma\varepsilon)} \, .
\end{equation}

Having an explicit relation between conserved and primitive quantities
has an impact on the efficiency of the numerical code, as it
eliminates an iterative process that is required, at least once per
each spacetime point. It is however unavoidable for general, e.g.,
tabulated, EOS.

\section{Non-vacuum spherically symmetric spacetimes in the
Tamburino-Winicour formalism}
\label{sec:general-1d}

We consider here the general spherically symmetric spacetime, with
perfect fluid matter, following the formalism of
Tamburino-Winicour~\cite{tamburino}, with a minor extension to cover
the case of a worldtube immersed in matter.

The Einstein equations sufficient for obtaining the spacetime
development are grouped as
\begin{eqnarray}
\label{g1}
G_{vr} & = & \kappa T_{vr}, \\
\label{g2}
G_{rr} & = & \kappa T_{rr}, \\ 
\label{g3}
G_{vv}|_{\Gamma} & = & \kappa T_{vv}|_{\Gamma},
\end{eqnarray}
where the $v$ coordinate is defined by the level surfaces of a null
scalar (i.e., a scalar $v$ satisfying $\nabla^{\mu} v \nabla_{\mu} v =
0$). The $r$ coordinate is chosen to make the spheres of rotational
symmetry have area $4 \pi r^2$. The $x^{2},x^{3}$ coordinates in this
geometry are simply taken to be the angular coordinates
$(\theta,\phi)$ propagated along the generators of the null
hypersurface, i.e., they parameterize the different light rays on the
null cone. The first two equations contain only radial derivatives
and are to be integrated along the null surface. The last
equation~(\ref{g3}) is a {\em conservation} condition to be imposed on
the world-tube $\Gamma$ (see
Fig.~\ref{fig:cones}). Equation~(\ref{g2}) may be substituted for by
the equivalent expression $g^{ab}R_{ab}=8\pi
g^{ab}(T_{ab}-g_{ab}T/2)$, where the indices $(a,b)$ run over the
remaining coordinates $x^{2},x^{3}$. To proceed with the integration,
an additional choice of gauge must be made {\em on the
worldtube}. This condition fixes the only remaining freedom in the
coordinate system, namely the rate of flow of coordinate time at the
world-tube.

We present here the explicit expressions we use, including the matter
terms. Adopting the Bondi-Sachs form of the metric element,
\begin{equation}
\label{eq:bondi-sachs}
ds^2 = - \frac{e^{2\beta}V}{r} dv^2 
+ 2 e^{2\beta} dv dr + r^2 (d\theta^2 + \sin\theta^2 d\phi^2) \, ,
\end{equation}
the geometry is completely described by the two functions $\beta(v,r)$
and $V(v,r)$ (we will also interchangeably use the variable $W = V -
r$).

The $\beta$ and $V$ hypersurface equations are given by
\begin{eqnarray}
\label{eq:beta}
\beta_{,r} & = & 2 \pi r T_{rr} \, ,\\
\label{eq:V}
V_{,r} & = & e^{2\beta} + 4 \pi r V T_{rr} + 8 \pi r^2 T_{rv} \, ,
\end{eqnarray}
the latter being equivalent (modulo the four-velocity normalization
condition) to
\begin{equation}
V_{,r} = e^{2\beta} (1 - 4 \pi r^2 (g^{AB}T_{AB} - T)) \, .
\end{equation}
\noindent
The comma in the above equations indicates, as usual, partial
differentiation.

Boundary conditions $(\beta(v)_{\Gamma},V(v)_{\Gamma})$ for the radial
integrations are provided by the equation~(\ref{g3}) which explicitly
reads,
\begin{equation}
\frac{V \beta_{,r}}{r} + \beta_{,v} - \frac{V_{,v}}{2V} 
- \frac{V_{,r}}{2r} + \frac{1}{2r} e^{2\beta} - 4 \pi \frac{r^2}{V} T_{vv} 
= 0 \, ,
\end{equation}
with the adoption of a suitable gauge condition. For each choice one
obtains a pair of ODE's along the worldtube. For example, with the
choices
\begin{eqnarray}
\label{worldtube1}
V_{,v} & = & - 8 \pi r^2 T_{vv} \, ,\\ \nonumber
V_{,v} & = & 0 \, ,\\  \nonumber
(V e^{2 \beta})_{,v} & = & 0 \, , 
\end{eqnarray}
one obtains
\begin{eqnarray}
\label{worldtube2}
\beta_{,v} & = & 4 \pi r T_{vr} \, ,\\ \nonumber
\beta_{,v} & = & 4 \pi r T_{vr} + 4 \pi r^2  T_{vv} / V_{0} \, ,\\ \nonumber
\beta_{,v} & = & 2 \pi r T_{vr} - 2 \pi e^{2\beta}  T_{vv} / g_{00} \, ,
\end{eqnarray}
respectively, where $V_{0}$ and $g_{00}$ denote the integration
constants at $v=0$.  All the above conditions are equivalent in
vacuum, and since our present computations place the worldtube in very
low density regions, we do not analyze the issue further.

The hydrodynamic equations reduce, within our symmetry assumptions, to:
\begin{eqnarray}
\label{eq:Dr}
D_{,v} + F^{r0}_{,r} & = & 
- (\ln{\cal V})_{,v} D -  (\ln{\cal V})_{,r} F^{r0} \, ,\\
\label{eq:Sr}
S^{r}_{,v} + F^{r1}_{,r} & = & 
- (\ln{\cal V})_{,v} S^{r} -  (\ln{\cal V})_{,r} F^{r1} 
- \Gamma^{r}_{\mu\nu} T^{\mu\nu} \, ,\\
\label{eq:Er}
E_{,v} + F^{r4}_{,r} & = & 
- (\ln{\cal V})_{,v} E -  (\ln{\cal V})_{,r} F^{r4} 
- \Gamma^{v}_{\mu\nu} T^{\mu\nu} \, ,
\end{eqnarray}
where ${\cal V}=\sqrt{-g}=r^{2}\sin\theta e^{2\beta}$ is the four
dimensional volume element. The precise form of the flux terms is
obtained with direct use of the general formulae~(\ref{eq:fluxes}) and
the Christoffel symbols $\Gamma^{v}_{\mu\nu}$ are derived explicitly
for metric~(\ref{eq:bondi-sachs}).

In summary, the initial value problem consists of
equations~(\ref{eq:beta},\ref{eq:V},\ref{worldtube1}-\ref{eq:Er})
together with initial and boundary data for the fluid variables
$(\rho,\varepsilon,u^{r})$ on the initial slice $\Sigma_0$ (at time
$v_0$) and the metric values of $\beta(v_0)_{\Gamma}$ and
$V(v_0)_{\Gamma}$ at the woldtube.  Those equations and initial data
are sufficient for obtaining the spacetime in a domain to the future
of the initial hypersurface, which is radially bounded by the
worldtube.

\subsection{Stationary configurations: Tolman-Oppenheimer-Volkoff 
solutions along the null cone}

We present here the equations describing a spherically symmetric
equilibrium configuration in null coordinates.
Such solutions constitute an excellent test-bed for consistency and
accuracy checks of our algorithms. For the simplest derivation, it will
be advantageous to use a slightly different form of the metric
element. With the redefinition $Y=V e^{-2\beta}$, it reads,
\begin{equation}
ds^2 = - \frac{e^{4\beta} Y}{r} dv^2 
+ 2 e^{2\beta} dv dr + r^2 (d\theta^2 + \sin\theta^2 d\phi^2) \, .
\end{equation}
In analogy with the spacelike foliated case, the stationarity in all
metric and fluid variables reduces the number of non-trivial equations
to the following coupled pair,
\begin{eqnarray}
p_{,r} &=& \left(\frac{1}{2r} - \frac{1}{2Y} (1+8\pi r^{2} p)\right) 
\rho h  \, , \label{mom-tov}
\\
Y_{,r} &=& 1 + 8 \pi r^2 (p - \rho h) \, ,
\label{mass-tov}
\end{eqnarray}
where it is easily recognized that equation~(\ref{mom-tov}) is the
radial momentum balance equation~(\ref{eq:Sr}). This system of
equations is solved as a system of ODE's.

\begin{figure}[tbh]
\centerline{\psfig{file=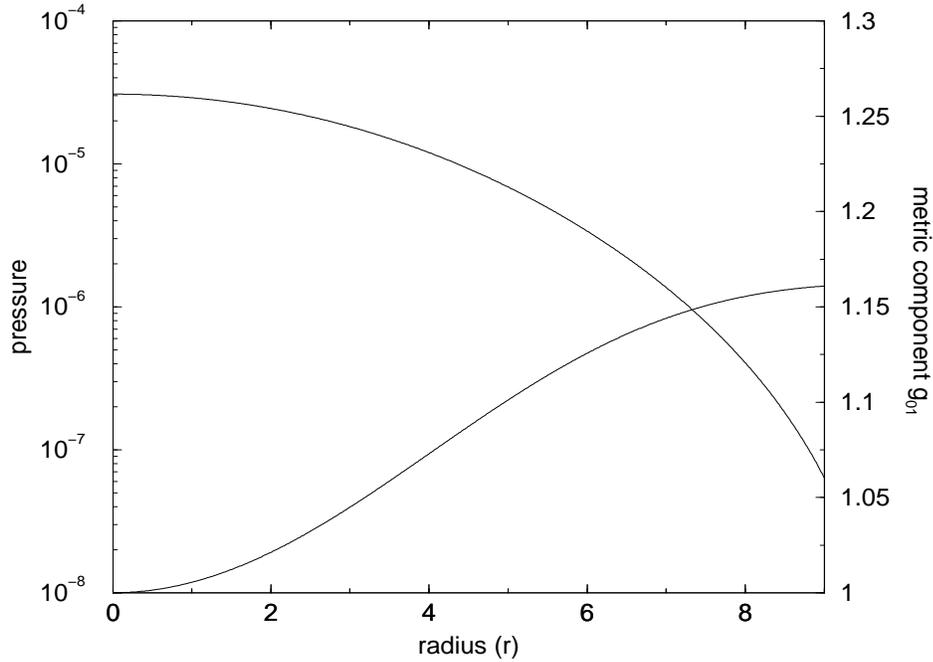,width=5.0in,height=4.0in}}
\caption{The pressure and $g_{01}$ metric component for a
representative TOV solution along the null cone (for $\Gamma=5/3$,
$K=4.349$, central density $\rho_{c}=8.1e-4$). A polytropic EOS is
assumed, i.e., $p=K\rho^{\Gamma}$.}
\label{fig:TOV-sols}
\end{figure}

In the special case of a ``stellar configuration'', solutions are
obtained by starting, e.g., with initial conditions $p=p_{c}, Y=0$ at
the center of the star and integrating outwards until the pressure
crosses the zero level~\cite{TOV-comment}. Following that recipe, the
direct integration of Eq.~(\ref{eq:beta}), with the boundary condition
$\beta=0$ at the origin, completes the metric element. Fields of a
representative solution, which we use in the next section for code
testing are presented in Fig.~\ref{fig:TOV-sols}.

\subsubsection{Constant density star}

In special cases the TOV equations can be integrated explicitly in
terms of simple functions. A widely known example is the {\em constant
density} star ($\rho_{0}$), derived here in ingoing null coordinates,
\begin{eqnarray}
\beta & = & \frac{1}{2} \ln{ \frac{3 A - B}{2 B} } \, , \\
    V & = & \frac{r B}{2} (3A - B) \, , \\
    p & = & \rho_{0} \frac{B-A}{3A-B} \, ,
\end{eqnarray}
where $A=\sqrt{1-2M/R}$, $B=\sqrt{1-2Mr^2/R^3}$ and $M=4\pi\rho_{0}R^3$,
with $R$ being the radius of the star.

\section{Algorithms and Tests}
\label{sec:alg-tests}

The general formalism outlined in the previous two sections can form
the basis for a variety of numerical approaches. Concerning the
GRH equations, in the original Wilson's scheme~\cite{wilson72}, a
combination of finite-difference upwind techniques with artificial
viscosity terms were used to damp spurious oscillations, extending the
classic treatment of shocks introduced by von Neumann (see,
e.g.,~\cite{rm67}) into the relativistic regime.  Artificial viscosity
based methods, though, were later shown to fail at the threshold of
the ultra-relativistic regime, once the Lorenz factor exceeds a value
of 2. Explicit HRSC codes, following the so-called ``Godunov
approach", appeared as a much more solid alternative.

Since the early nineties, it has been gradually demonstrated (see,
e.g.,~\cite{betal97} and references therein), that methods exploiting
the hyperbolic character of the hydrodynamic equations are optimally
suited for accurate integrations, even well inside the
ultra-relativistic regime. As mentioned previously, these schemes are
commonly known as {\it high-resolution shock-capturing} schemes. In a
HRSC scheme, the knowledge of the characteristic fields (eigenvalues)
of the equations, together with the corresponding eigenvectors, allows
for accurate integrations, by means of either exact or approximate
Riemann solvers, along the fluid characteristics.  These solvers,
which constitute the kernel of our numerical algorithm, compute, at
every interface of the numerical grid, the solution of local Riemann
problems (i.e., the simplest initial value problem with discontinuous
initial data). Hence, HRSC schemes automatically guarantee that
physical discontinuities appearing in the solution, e.g., shock waves,
are treated consistently (the {\it shock-capturing} property).  HRSC
schemes are also known for giving stable and sharp discrete shock
profiles. They have also a high order of accuracy, typically second
order or more, in smooth parts of the solution.

We proceed now to describe the highlights of our implementation.  The
grid structure is, in summary, as follows: An equidistant radial grid
$r_{i}$, with spacing $\Delta r$, denotes the location of cell centers
on which the conserved variables and metric element components
reside. The interface locations $r_{i}^{I}=r_{i}-\Delta r/2$ are used
for the reconstruction of variables and the solution of the Riemann
problems. Appropriate number of {\em ghost-zones} are added on the
boundaries of the grid to allow imposition of boundary conditions.  In
particular, the worldtube
conditions~(\ref{worldtube1}-\ref{worldtube2}) are discretized in time
along the first ghost-zone at $r_{N+1}$, where $r_{N}$ denotes the
last cell center inside the domain. The timestep is variable and
consistently computed to satisfy the Courant condition for the
hydrodynamic equations. The geometric equations, being ODE's in
spherical symmetry, do not impose any timestep restriction.

Our numerical implementation of the coupled system is broadly based on
the notion of operator splitting, i.e., the integration of the initial
value problem in steps, by successive applications of split components
of the overall evolution operator. We describe here, schematically,
the procedure. The set of equations, comprised of the three
hydrodynamical equations in spherical symmetry, and the hypersurface
equations governing the geometry, are written as
\begin{eqnarray}
\partial_{v} {\bf U} &+& \partial_{r} {\bf F}({\bf w},{\bf G}) = 0 \, , \\
\partial_{v} {\bf U} & & \hspace{1.75cm} =  {\bf S}_{U}({\bf w},{\bf G}) \, ,\\
& & \hspace{-0.17cm} {\bf A}( {\bf U}, {\bf w}, {\bf G}) = 0 \, ,\\
& & \hspace{0.0cm} \partial_{r} {\bf G} \hspace{0.98cm}
=  {\bf S}_{G}({\bf w},{\bf G}) \, .
\label{gr-ode}
\end{eqnarray}
The ordering of the equations reflects the sequence in which they are
solved in the code. We denote the metric variables collectively as
${\bf G}$, while ${\bf A}$ stands for the set of algebraic relations
connecting conserved and primitive variables (with the mediation of
the metric). The velocity normalization condition is included with
this set.

The first step in the solution process involves an advection of the
conserved variables ${\bf U}$, from the initial data hypersurface
$\Sigma_{0}$ to the hypersurface $\Sigma_{\Delta v}$. The computation
of the fluxes {\bf F} uses the metric as it is given in $\Sigma_{0}$,
along with the primitive variables that are assumed to have been
computed on $\Sigma_{0}$ in the previous iteration. The advection may
be performed by any modern numerical scheme for the propagation of
non-linear waves, i.e., by any HRSC scheme. These schemes,
being written in conservation form, are particularly well suited for this
purpose. Hence, all conserved quantities in the differential equations
are also conserved in their finite-differenced versions. More precisely,
the time update of 
\begin{eqnarray}
\partial_v {\bf U} + \partial_r {\bf F} = 0 \, ,
\end{eqnarray}
is done according to the following algorithm:
\begin{eqnarray}
    {\bf U}_{i}^{n+1}={\bf U}_{i}^{n}-\frac{\Delta v}{\Delta r}
    (\widehat{{\bf F}}_{i+1/2}-\widehat{{\bf F}}_{i-1/2}) \, .
\end{eqnarray}
\noindent
The index $n$ represents the time level, while the time discretization
interval is indicated by $\Delta v$.  The ``hat" in the fluxes is used
to denote the so-called numerical fluxes which, in a HRSC scheme, are
computed according to some generic flux-formula, of the following
functional form:
\begin{eqnarray}
  \widehat{{\bf F}}_{i\pm{1\over 2}} = \frac{1}{2}
          \left( {\bf F}({\bf U}_{i\pm{1\over 2}}^{L})  +
                 {\bf F}({\bf U}_{i\pm{1\over 2}}^{R}) -
          \sum_{\alpha = 1}^{p} \mid \widetilde{\lambda}_{\alpha}\mid
          \Delta \widetilde {\omega}_{\alpha}
          \widetilde {r}_{\alpha} \right) \, .
\end{eqnarray}
\noindent
Notice that the numerical flux is computed at cell interfaces
($i\pm1/2$).  Indices $L$ and $R$ indicate the left and right sides of
a given interface. The sum extends to $p$, the total number of
equations. Finally, quantities $\lambda$, $\Delta\omega$ and $r$
denote the eigenvalues, the jump of the characteristic variables and
the eigenvectors, respectively, computed at the cell interfaces
according to some suitable average of the state vector variables.

In our code, the numerical integration of the hydrodynamic equations
can be performed using two different approximate Riemann
solvers. These are the Roe solver~\cite{roe81}, widely employed in
fluid dynamic simulations, with arithmetically averaged states (for
the use of the Roe mean in a relativistic Roe solver see~\cite{em95})
and the Marquina solver, recently proposed in~\cite{donat96} (see
also~\cite{donat98}).

After the update of the transport terms the fluid variables are
subsequently corrected for the effect of the source terms. Any stable
ODE integrator is usable here. Our choice is a second order
Runge-Kutta method. A point of interest here is that the source terms
${\bf S}_{U}$ depend on both fluid and metric variables, and, in
particular, on time derivatives of the latter. This implies that a
second order capturing of the effect of those derivatives requires
storing an additional time level of metric variables.

The geometry equations comprise a system of radial ODE's, in which the
right-hand-side depends on both the metric and the stress-energy
tensor, a system to be solved on the hypersurface $\Sigma_{\Delta
v}$. Initial conditions are provided on the world-tube.  Hence, the
integration cannot proceed without the recovery of the primitive
variables on $\Sigma_{\Delta v}$, using the expressions given in
section~\ref{primitives}. It can be seen that those expressions
involve the metric (which is to be solved for). Furthermore, one must
worry about the preservation of the velocity normalization condition,
which strongly depends on the metric at a given point. This
differential-algebraic entangling of metric and fluid variables,
generic to any attempt to solve general fluid spacetimes using
conservative formulations, is approached in the following way:

The ODE integrator for equation~(\ref{gr-ode}) is chosen to belong to
the implicit class. Our present choice is the second order Euler
method. The metric variables in the new radial location $r_{i+1}$ are
obtained iteratively ($k$ denotes the iteration index). Inside the
$k$-iteration loop the intermediate metric variables ${\bf
G}_{i+1}^{k}$ are used to obtain intermediate primitive variables
${\bf w}_{i+1}^{k}$, which then leads to updated values for the source
term ${\bf S}_{G,i+1}^{k}$. In addition, the zeroth component of the
velocity vector is adjusted so that the primitive velocity field
$u^{\mu}$ is normalized. This process can be described as
\begin{eqnarray}
& & {\bf A}( {\bf U}_{i+1}, {\bf w}_{i+1}^{k}, {\bf G}_{i+1}^{k}) = 0 \, ,\\
& & {\bf G}_{i+1}^{k} - {\bf G}_{i} 
=  \frac{\Delta r}{2} ({\bf S}_{G,i+1}^{k} + {\bf S}_{G,i}) \, ,
\end{eqnarray}
(the time level index has been suppressed here for clarity). In practice
the number of required iterations is about four or five.  The
completion of this procedure furnishes simultaneously the metric and
the primitive variables at the new radial location, at the new time
level.

\subsection{The Riemann problem on a null Minkowski slice}

We proceed now to establish the feasibility of our proposal by
performing numerical integrations of the GRH equations on null
surfaces. Using the characteristic fields derived previously, we show
here with two numerical examples the capabilities of existing
HRSC schemes to integrate discontinuous solutions described in
characteristic spacetime foliations. The initial data are given on a
null slice (advanced or retarded time) of Minkowski space, and the
evolution is compared to the suitably transformed exact solution.

The shock tube (a particular case of the Riemann problem) is one of
the standard tests to calibrate numerical schemes in classical fluid
dynamics. The initial setup of this experiment consists of a zero
velocity fluid having two different thermodynamical states on either
side of an interface. When this interface is removed, the fluid
evolves in such a way that four constant states occur.  Each state is
separated by one of three elementary waves: a shock wave, a contact
discontinuity and a rarefaction wave.  This time-dependent problem has
an exact solution~\cite{godunov}, to which the numerical integration
can be compared. In addition, it provides a severe test of the
shock-capturing properties of any numerical scheme.  In recent years,
the shock tube problem began being used as a test of (special)
relativistic hydrodynamical codes (see, e.g.,~\cite{donat98} and
references therein). The analytic solution of the Riemann problem,
also available in relativistic hydrodynamics since~\cite{mm1}, allows
for a rapid and unambiguous comparison with the numerical evolutions.

For our numerical demonstrations we consider two different initial
setups.  In case 1 the initial state of the fluid is specified by
$p_L=13.3$, $\rho_L=10$ on the left side of the interface and $p_R=0$,
$\rho_R=1$ on the right side. For numerical reasons, the pressure of
the right state is set to a small finite value ($p_R=0.66\cdot
10^{-6}$). Case 2 is as case 1 but with the left and right states
reversed.  We use a perfect fluid EOS with $\Gamma=5/3$.

\subsubsection{The advanced time case}

\begin{figure}[tbh]
\centerline{\psfig{figure=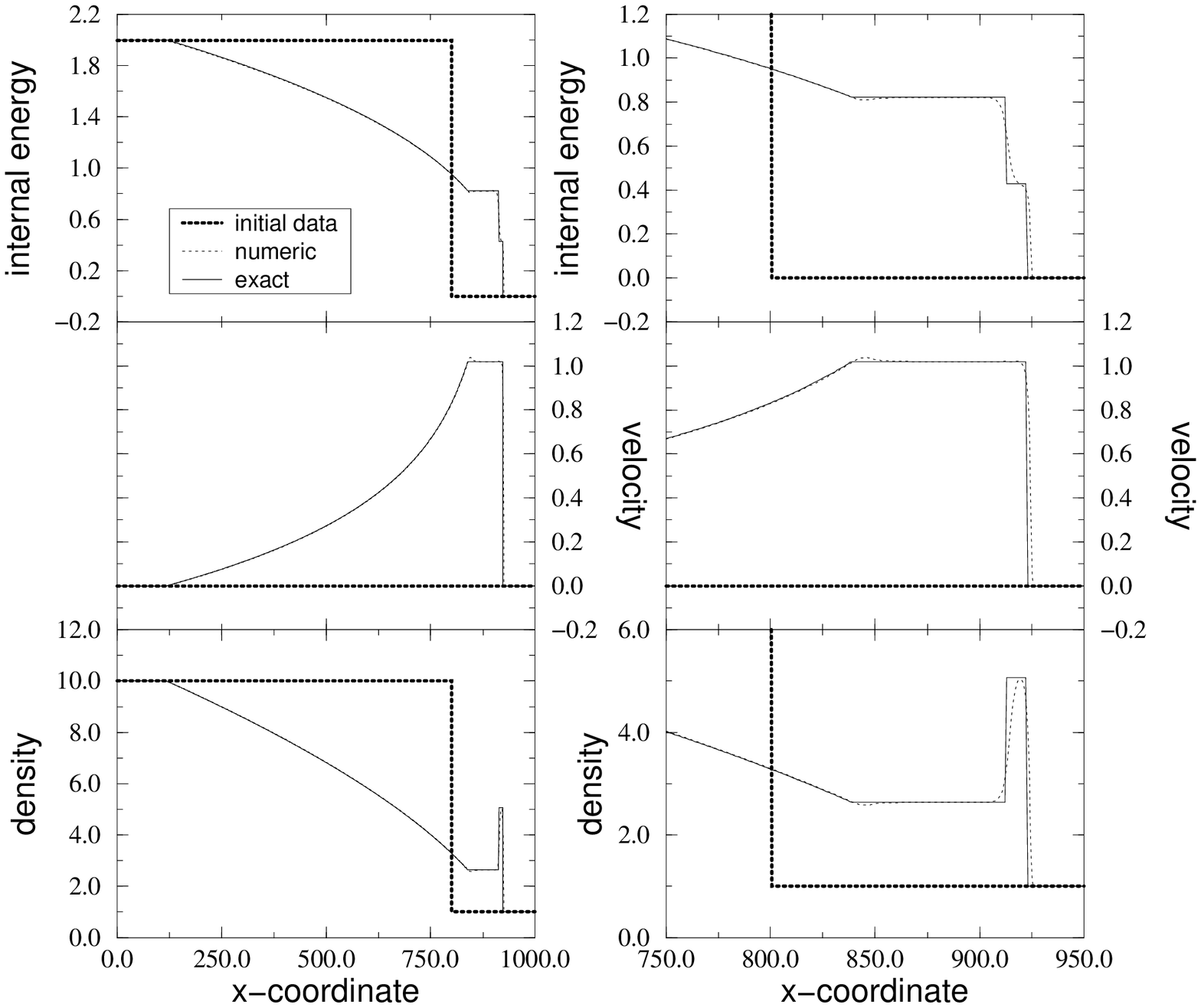,width=6.0in,height=8.0in}}
\caption{ The {\it outgoing} shock tube problem (case 1). Exact versus
numerical solution at an advanced time $v=270$. The domain extends
from $x=0$ to $x=1000$ and a grid of 2000 zones is used ($\Delta
x=0.5$). From top to bottom we plot the internal energy, velocity and
density. The right panels show a zoomed view of the left panels
focusing on the most interesting region. The thick dashed line shows
the location of the initial discontinuity. The solid line is the
exact solution and the dotted line is the numerically computed one.
Those features of the solution moving to the right, i.e., the shock
wave and contact discontinuity, are less developed than those moving
to the left, i.e., the rarefaction wave.  }
\label{fig:shock1}
\end{figure}

In Fig.~\ref{fig:shock1} we plot the results for case 1. The
left panels show the whole domain, with the $x$-coordinate ranging
from 0 to 1000. The right panels show a zoomed up view of the most
interesting domain ($x\in[750,950]$). We use a numerical grid of 2000
zones and, hence, a rather coarse spatial resolution ($\Delta x=0.5$).

>From top to bottom, Fig.~\ref{fig:shock1} displays the internal energy
($\varepsilon$), the velocity ($u^x$) and the density ($\rho$).  The
thick dotted line represents the initial discontinuity.  The 
solid line shows the exact solution after an advanced time $v=270$ and
the dotted line indicates the numerical solution at this same time. In
order to compute the exact solution in our new coordinates we have
followed the procedure outlined in~\cite{mm1} applying the appropriate
time transformation, i.e., $t=v-x$.  Clearly, the agreement between
the exact (solid lines) and numeric solution is remarkable. The
solution is characterized by an (outgoing) shock wave (moving to the
right) and an (ingoing) rarefaction wave (moving to the left).  One
important point to notice is that features of the solution moving
towards the left are more developed than those moving to the
right. This is visible in the respective locations of the head of the
rarefaction and the shock wave with respect to the initial
discontinuity. The appearance of the rarefaction wave differs from the
standard ``spacelike" one (see, e.g., Fig.~2 of~\cite{donat98}), due
to the intervening time transformation. The internal energy and the
density show an additional elementary wave, a contact discontinuity
between the shock and rarefaction waves. Across the contact
discontinuity pressure and velocity are constant, while the density
exhibits a jump.

In the right panels we note that the largest discrepancies are found
at the tail of the rarefaction, in the form of an ``undershooting" in
density and internal energy and an ``overshooting" in velocity. This
is a typical feature of the numerical solution of the shock tube
problem and it is not related to the null coordinates used here. We
note that the constant state between the shock wave and the contact
discontinuity is well resolved despite the coarse resolution used. For
comparison purposes, the interested reader is referred
to~\cite{donat98}, where similar results were obtained (in a spacelike
approach) employing a grid resolution 200 times finer.

\subsubsection{The retarded time case}

\begin{figure}[tbh]
\centerline{\psfig{figure=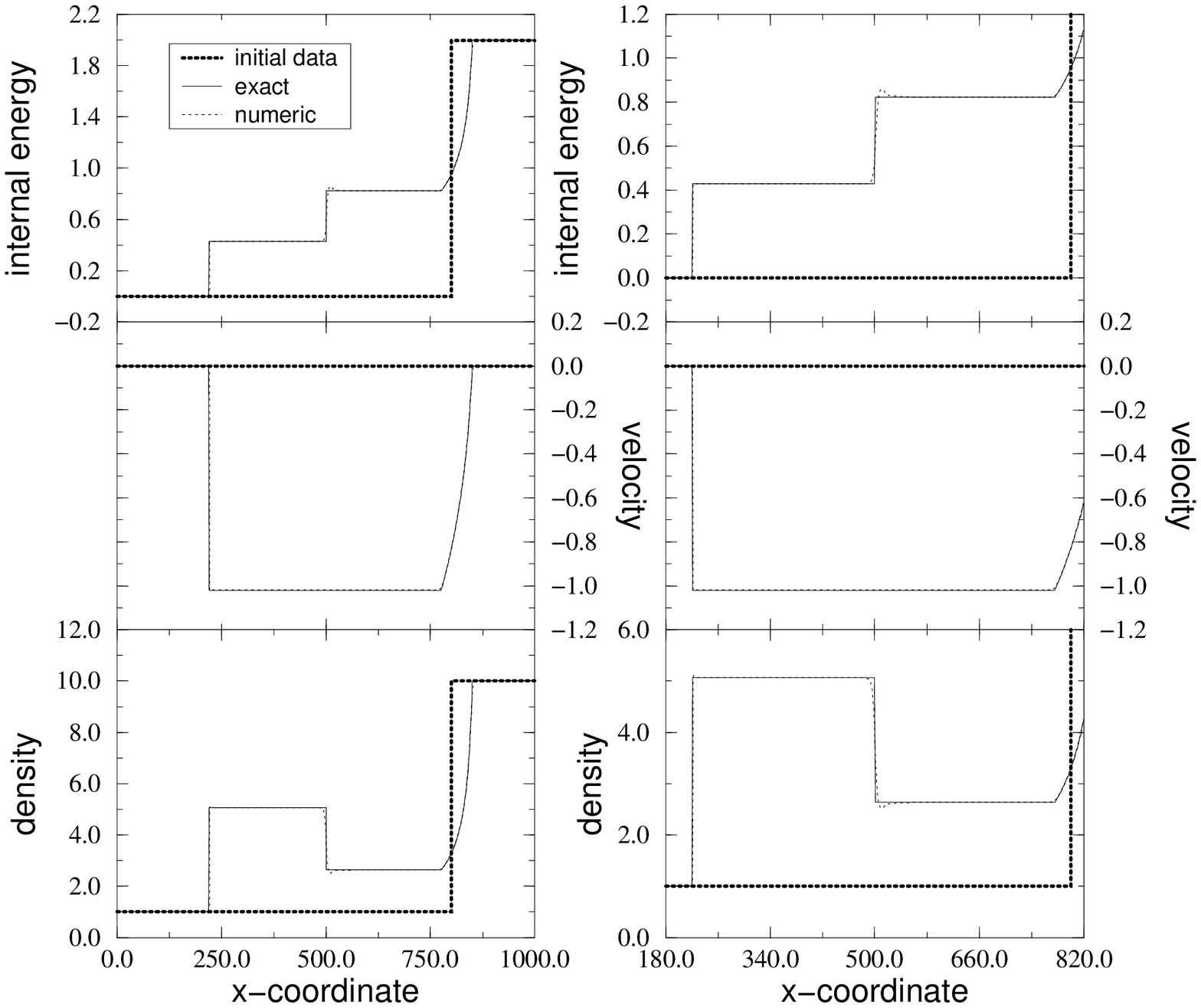,width=6.0in,height=8.0in}}
\caption{ The {\it ingoing} shock tube problem (case 2). Exact versus
numerical solution at an advanced time $v=120$. The domain extends
from $x=0$ to $x=1000$ and a grid of 2000 zones is used ($\Delta
x=0.5$). From top to bottom we plot the internal energy, velocity and
density. The right panels show a zoomed view of the left panels
focusing on the most interesting region. The thick dashed line shows
the location of the initial discontinuity. The solid line is the
exact solution and the dotted line is the numerically computed one.  As
in Fig.~\ref{fig:shock1}, those features of the solution moving to the
right, i.e., the rarefaction wave, are less developed than those
moving to the left, i.e., the shock wave and contact discontinuity.  }
\label{fig:shock2}
\end{figure}

In Fig.~\ref{fig:shock2} the results for the ``mirror" version of the
shock tube problem 1 are plotted. As in Fig.~\ref{fig:shock1}, the
left panels show the whole domain, with the $x$-coordinate ranging
from 0 to 1000 whereas the right panels show a closed-up view of the
most interesting region ($x\in[180,820]$). We use the same grid
resolution as in case 1. The initial data are now evolved up to a
final time $v=120$.  Again, those features of the solution moving
towards the left are more developed than those moving towards the
right.  The ``tower" in the density is now considerably wider than in
case 1 as well as the intermediate constant state in the internal
energy between the shock and the contact discontinuity (almost
not noticeable in case 1). This constant state is much more accurately
resolved than in case 1 despite the use of the same resolution. The
two corners of the rarefaction wave are also very well resolved. The
major numerical deviations from the exact solution appear now at the
post contact discontinuity region, with the density and the internal
energy being slightly under and over estimated there.

\begin{figure}[t]
\centerline{\psfig{figure=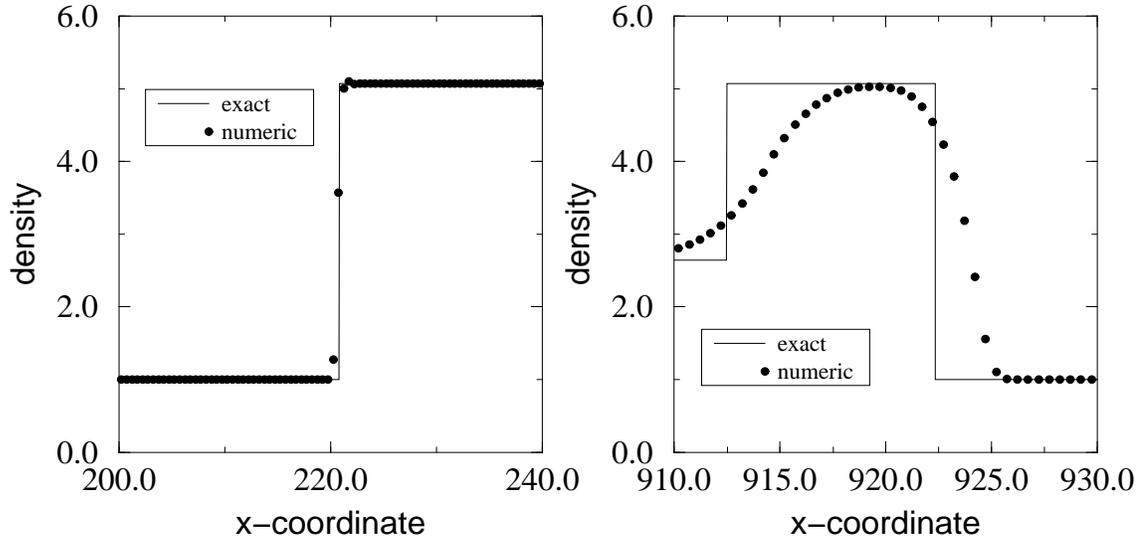,width=6.0in,height=4.3in}}
\caption{ Capturing the shock wave. This figure shows the different
ways the shock wave is resolved in the shock tube problems 1 (ingoing;
left) and 2 (outgoing; right). In spite of the fact that the grid
resolution is the same in both simulations, in case 2 the shock wave
is spread out in a large number of cells whereas in case 1 it is
sharply captured in two zones (out of a total number of 2000 zones).
}
\label{fig:shock3}
\end{figure}

We should comment on the different ways in which the shock is captured
in cases 1 and 2. Whereas in case 1 the shock is spread out in a
number of grid zones ($\approx 10$), in case 2 it is sharply resolved
in 2-3 cells (out of the 2000 zones used). This is shown in
Fig.~\ref{fig:shock3}. Those differences ultimately derive from the
fact that a retarded (or advanced) time formulation of a Riemann
problem (assuming, for the discussion, the initial discontinuity
located at $x=0$) breaks the commutation between the operation of
reflection symmetry (between positive and negative values of $x$) and
that of time evolution. The significance of this effect in
applications involving shocks and its possible exploitation for
ultrarelativistic outflows will be studied elsewhere.

\begin{figure}[tbh]
\centerline{\psfig{figure=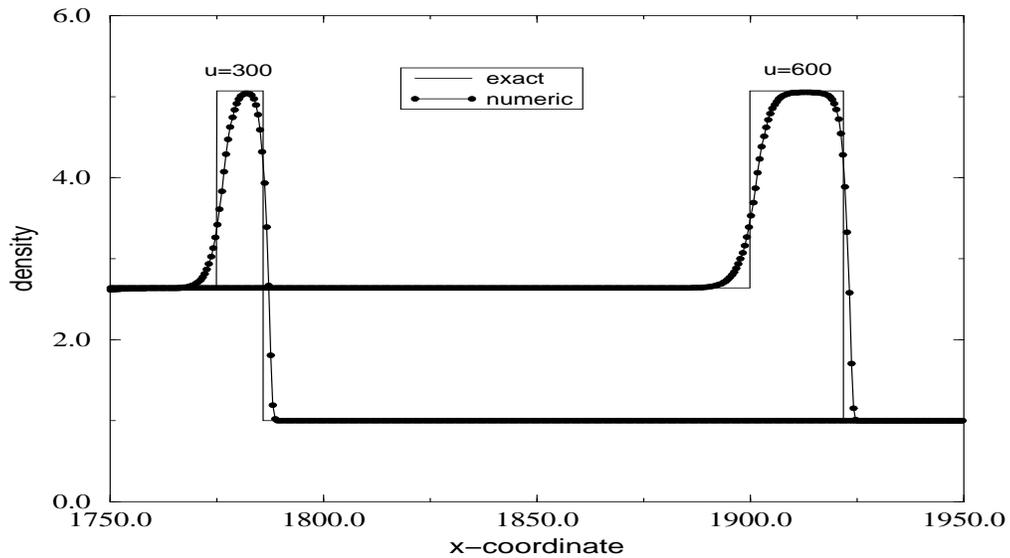,width=6.0in,height=3.4in}}
\caption{ Location of the outgoing shock in the shock tube problem 1
at two different evolution times, $v=300$ and $v=600$. The numerical
solution spreads the shock in a constant number of zones ($\approx
10$) from the very start and throughout the evolution.  }
\label{fig:shock4}
\end{figure}

We also include in Fig.~\ref{fig:shock4} the evolution of the outgoing
shock of case 1 at two different times, $v=300$ and $v=600$. For this
run we are using $4000$ zones with the $x$-coordinate ranging from 0
to 2000. Hence, the spatial resolution is again $\Delta x=0.5$.  The
initial discontinuity is placed at $x=1650$. The solution was left to
evolve for a time considerably longer than before in order to find out
if the numerical representation of the shock had more time to
accommodate itself into a sharper profile. From Fig.~\ref{fig:shock4}
we find this not to be the case. The initial spreading present in the
numerical solution remains constant in time. It is noteworthy to
mention the independence of the number of points in which the
shock is resolved from the total number of zones used.

\subsection{Convergence testing of the coupled code}

The TOV solutions presented before constitute a good test-bed for
confirming both the consistency and the accuracy of the algorithm.
The availability of semi-analytic solutions to the stationary problem
means that there is a host of {\em error functions} that can be
constructed by subtracting the evolved solution at a fixed total time
$t_{F}$ from the initial profile. As an example, using the exact density
profile $\rho_{E}$, the quantity
\begin{equation}
\|\rho-\rho_{E}\|_{2} = \sum_{i} (\rho_{i} - \rho_{E})^{2}
\end{equation}
measures the deviations produced by the numerical evolution.

\begin{figure}[tbh]
\centerline{\psfig{file=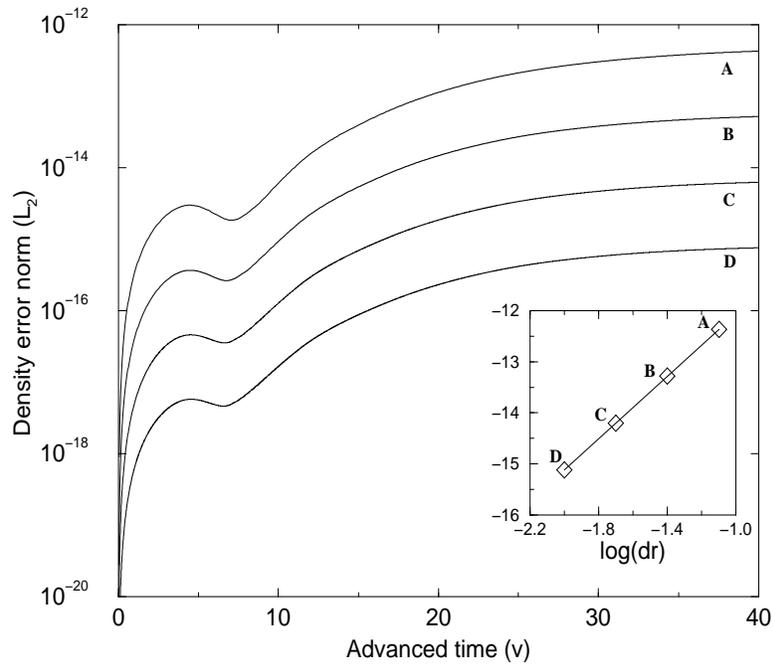,width=4.5in,height=4.0in}}
\caption{The $L_{2}$ norm of the density function (i.e., the
integrated squared difference of the numerical from the exact
solution) is plotted against time, for four successive doublings of
the grid size (curves A,B,C,D). The norm converges to third
order. This can be seen in the insert, where the final values of the
norm at $v=40$ are plotted against grid size (diamonds). The best
linear fit, represented by the line, has slope 3.05. This
implies that the local error is second order.}
\label{fig:conv}
\end{figure}

As our present implementations are geared towards black hole
spacetimes (with topology of $S^{2} \times R$ rather than regular
$R^{3}$ spacetimes), in order to benefit from this test-bed without
undue boundary complications, we consider only a {\em portion} of the
spacetime, excluding the domain around $r=0$. Initial data for the
fluid and the metric are obtained with the solution of the TOV
equations along the past null-cone of the center of symmetry. The
integration then proceeds in the regular fashion described above, but
restricted to a domain inside the star, e.g., in the sample shown in
Fig.~\ref{fig:TOV-sols} this extends between $r=2$ and $r=6$.  In the
plots given in Fig.~\ref{fig:conv} we consider the convergence of such
functions and their $L_{2}$ norms to zero, as the grid spacing is
reduced. The norm is found to converge to third order. This can be
seen in the insert of Fig.~\ref{fig:conv}, where the final values of
the norm at $v=40$ (about eight light crossing times) are plotted
against grid size. This implies that the local error is second order.

The constant density solution is obtained with the assumption of an
ad-hoc (and largely unphysical) equation of state and hence it is not
possible to evolve such data with the formalism developed here, but it
is straightforward to test the integration of the hypersurface
equations against the exact solution. We have confirmed that the
integrated metric agrees along the hypersurface with the exact value
to second order in the radial grid spacing.

\section{Spherical accretion onto a black hole}
\label{sec:accretion}

We proceed now in applying the formalism, in spherical symmetry, to
the problem of interaction of matter with a black hole. In recent
work~\cite{marsa96} the interaction of a scalar field with a black
hole was investigated, partly as a probe into the concept of
``singularity excision''~\cite{thornburg,seidel-suen,scheel}.
Following that same
concept in spirit, in a previous investigation, we studied fluid
interactions with the black hole geometry, {\em in the test-fluid
limit}. We dubbed stationary coordinates that are regular at the
horizon of an exact stationary black hole solution as {\it horizon
adapted coordinate systems}~\cite{HACS1}.  In those coordinates the
flow solution is smooth and regular at the black hole horizon.  The
steepness of the hydrodynamic quantities dominates the solution only
near the real singularity. This approach is now being applied in
successively more general (and astrophysically interesting) fluid
configurations~\cite{HACS2,HACS3}. We extend here this line of work by
first performing test-fluid computations in the spirit of~\cite{HACS1}
for the null Eddington-Finkelstein coordinate system. This is next
further generalized to account for self-gravitating accretion flows.

\subsection{The test fluid limit}

We study spherical accretion of a perfect fluid onto a (static) black
hole. The fluid is taken to have a sufficiently low density so that
during the accretion process the mass of the black hole remains
unchanged.  Stationary solutions to this idealized problem can be
computed exactly, up to algebraic equations. This was first derived
for Newtonian flows by Bondi~\cite{bondi}. The extension to general
relativity was due to Michel~\cite{michel}. The solution can easily be
re-derived for coordinates other than the original Schwarzschild
system employed by Michel (the details can be found
in~\cite{HACS1}). We use next this exact solution to quantify the
accuracy of our numerical integrations. The significance of the test
lies in its capturing of large curvature gradients near the black hole
(i.e., it is a strong field computation), which translates into the
existence of large source terms in the hydrodynamic equations.

We use ingoing Eddington-Finkelstein $(v,r,\theta,\phi)$ coordinates
for which the line element reads
\begin{eqnarray}
ds^2=-\left(1-\frac{2M}{r}\right)dv^2 + 2dvdr + r^2(d\theta^2+
\sin^2\theta d\phi^2) \, ,
\label{EF}
\end{eqnarray}
\noindent
where $M$ is the mass of the black hole. 
The expressions for the characteristic fields of the hydrodynamic
equations are then specialized using the above metric components.
We use a polytropic EOS with $\Gamma$ (the adiabatic exponent of the
gas) equal to $5/3$. The numerical domain extends from any given
non-zero radius inside the horizon, $r_{min}$, to some outer radius
$r_{max}$ (outside and far from the black hole horizon). In the
particular simulation reported here we choose $r_{min}=1.5M$ (inside
the black hole horizon, located at $r=2M$) and $r_{max}=30M$. We use a
uniform (equally spaced) numerical grid of 200 zones.

\begin{figure}[tbh]
\centerline{\psfig{figure=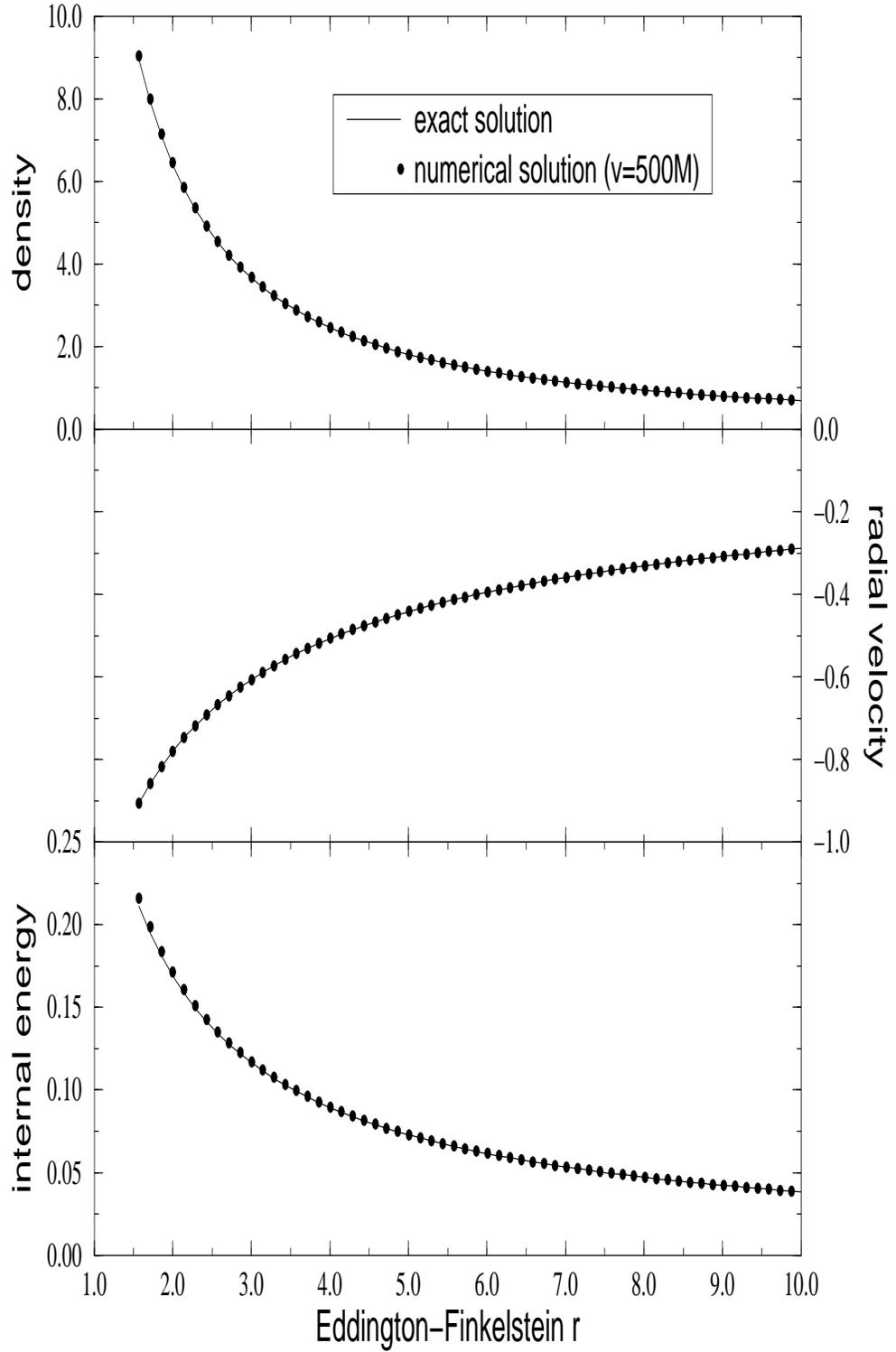,width=5.5in,height=8.0in}}
\caption{ Perfect fluid spherical accretion in ingoing
Eddington-Finkelstein coordinates (test fluid limit).  This figure
compares the exact (solid lines) and numerical (filled circles)
solutions for the hydrodynamical primitive variables, as a function of
the radial coordinate. The numerical solution is evolved up to a time
$v=500M$. From top to bottom we plot the density, velocity and
internal energy. The domain extends from $1.5M$ to $30M$ and a uniform
grid of 200 zones is used. Only the first $10M$ are shown.  }
\label{fig:accretion1}
\end{figure}

\begin{figure}[tbh]
\centerline{\psfig{figure=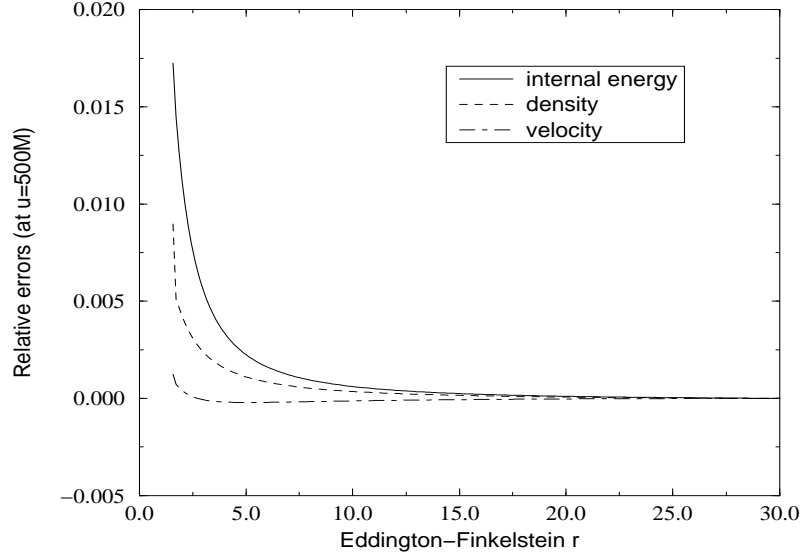,width=4.5in,height=3.4in}}
\caption{ Perfect fluid spherical accretion in ingoing
Eddington-Finkelstein coordinates (test fluid limit): relative errors
of the primitive variables. The maximum errors are less than 2\% for
the internal energy, 1\% for the density and 0.1\% for the
velocity. These numbers correspond to a simulation using 200 radial
zones in the interval $1.5M-30M$ and for an evolution up to $v=500M$
into the future.  }
\label{fig:accretion2}
\end{figure}

\begin{figure}[tbh]
\centerline{\psfig{figure=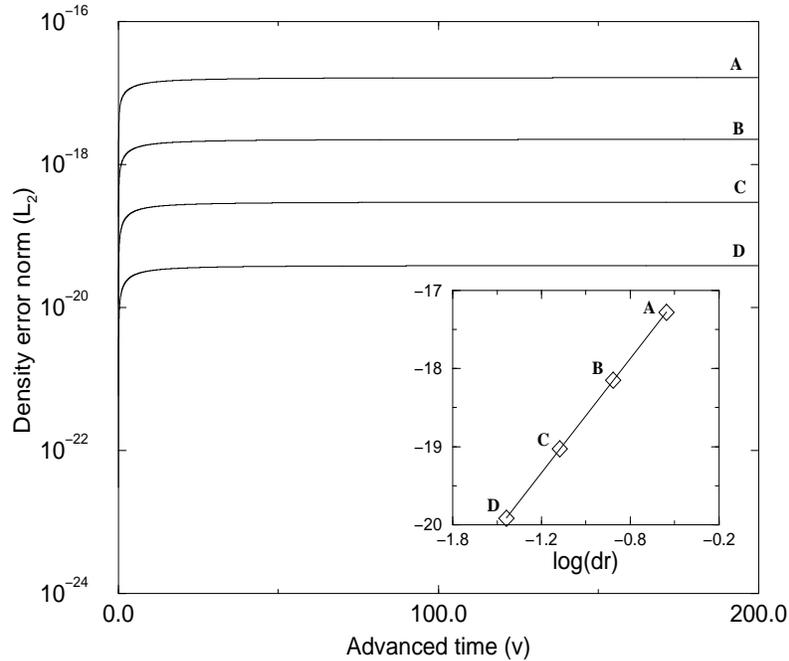,width=4.5in,height=4.0in}}
\caption{ Perfect fluid spherical accretion in ingoing
Eddington-Finkelstein coordinates (test fluid limit): converging
towards the stationary state. The four curves, labeled A,B,C,D,
present the time evolution of the density $L_{2}$ error norm (see
Fig.~\ref{fig:conv} for definition), for successive
doublings of the resolution. The insert shows the convergence of the
norm at final time ($v=200M$). The diamonds are measured values, the
line is a linear best fit with slope 2.9. Of interest in this plot is the
fast approach of the numerical solution to a steady state. }
\label{fig:fixed}
\end{figure}

The test proceeds as follows: We set the semi-analytic solution as
initial data throughout the domain, and then evolve these data,
maintaining the exact solution as a constant inflow boundary condition
at the outer boundary.  Representative results are summarized in
Figs.~\ref{fig:accretion1}-\ref{fig:fixed}.

In Figure~\ref{fig:accretion1} we display only the solution up to a
radius of $10M$, in order to focus on the most interesting, strong
field region, around the black hole horizon.  The figure displays the
primitive variables, ($\rho,u^r,\varepsilon$), as functions of the
ingoing Eddington-Finkelstein radial coordinate. These figures show
the capturing of the steady-state spherical accretion solution. The
solid lines represent the exact solution, while the filled circles
indicate the numerical one. The latter has been evolved up to a time
$v=500M$.  The agreement between the exact and numerical solutions is
very good for all fields. This is more clearly seen in
Fig.~\ref{fig:accretion2} where we plot the relative errors of the
primitive variables. The largest errors appear at the innermost
zones. The maximum error never exceeds 2\%, for the internal energy,
or 1\%, for the density, even with the relatively coarse grid of 200
points. The quality of the results for the velocity is excellent, with
the maximum relative error being about 0.1\%.

In Fig.~\ref{fig:fixed} the convergence properties are captured more
quantitatively. The squared difference of the density from the exact
result, integrated over the entire domain (the $L_{2}$ norm), is
plotted as a function of advanced time, for a total time of 200.
After an initial rise, the error settles into a final state, the level
of which is converging to third order with the radial grid size.

We note in passing that no secular instabilities of any kind arise
during the computation, for total number of iterations of the order of
$10^5$. Besides the intrinsic value of the test as an exact strong
field solution, an important practical aspect merits mentioning here:
very low density spherical inflow solutions constitute a good {\em
background} flow, useful in computations where the primary dynamics of
interest involves high density concentrations around the black hole.

\subsection{Accretion of a self-gravitating perfect fluid}

In our last numerical demonstration we investigate the accretion of a
self-gravitating perfect fluid onto a non-rotating black hole.  The
setup of the simulation is as follows: Boundary values corresponding
to a black hole of given mass $M_{F}$ are specified at the world-tube,
located at $r_{B}$. The interior to that radius is filled first with
low density data which do not interact with the black hole geometry in
the timescales of interest.  The dynamically interesting fluid
component, typically a high density distribution (compared to the
background) of compact support with sufficiently strong self-gravity,
is added next. The data to be specified are $(\rho, \epsilon, u^{r})$
and are, in short, completely arbitrary, with one exception: For the
setup to correspond to a spacetime with a trapped region, i.e., with a
horizon, the added fluid mass should not exceed $M_{F}$. The velocity
profile is borrowed from the Bondi accretion solution and, hence,
corresponds to an inwards monotonically increasing velocity. The
initial data for the density are specified with explicit profiles,
e.g., flat or Gaussian radial distributions.  In our simulations we
consider a Gaussian spherical shell surrounding the central black
hole, with density parameterized according to
\begin{eqnarray}
\rho=\rho_b + \rho_{m}e^{-\sigma(r-r_c)^2}
\end{eqnarray}
\noindent
where $\rho_b$ is the background density. The rest of parameters take
the values: $\rho_m=10^{-4}$, $\sigma=0.1$ and $r_c=6M$. The grid
extends from $r_{min}=1.1M$ to $r_{max}=20M$.  Finally, the internal
energy is obtained assuming an initially isentropic distribution of
pressure $p= K \rho^{\Gamma}$ (which would be valid at later times
only for equilibrium configurations).

\begin{figure}[t]
\centerline{\psfig{figure=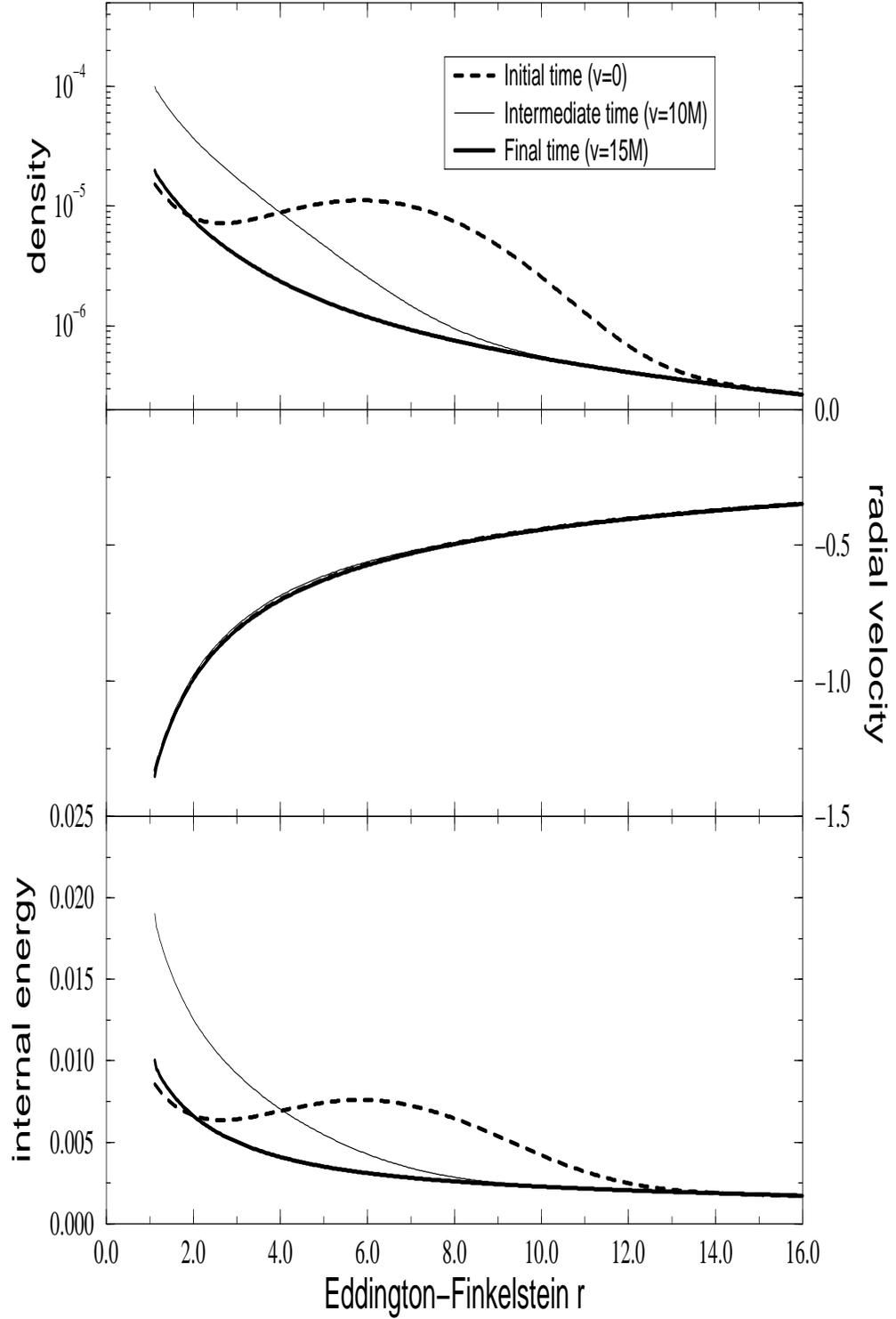,width=5.5in,height=8.0in}}
\caption{{Spherical accretion of a self-gravitating perfect fluid:
evolution of the primitive variables. Ingoing Eddington-Finkelstein
coordinates are used, in a grid of 500 zones spanning the radial
interval between $1.1M$ and $20M$. Three different times of the
evolution are shown: $v=0$ (dotted thick line), $v=10M$ (solid thin
line) and $v=15M$ (solid thick line).  The spherical shell, centered
at $r=6M$, is radially advected towards the hole. A final
quasi-stationary (Bondi) solution is achieved.  }}
\label{fig:sga1}
\end{figure}

\begin{figure}
\centerline{\psfig{figure=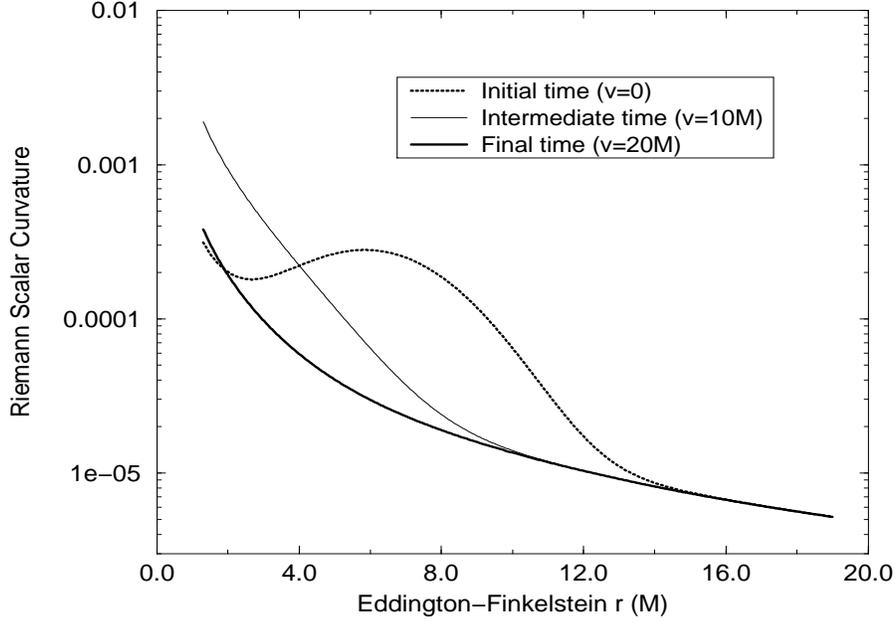,width=5.0in,height=3.8in}}
\caption{{ Spherical accretion of a self-gravitating perfect fluid:
evolution of the Riemann scalar curvature for the same simulation of
Fig.~\ref{fig:sga1}. Notice the non-vanishing self-gravity of the
initial distribution, demonstrated by a curvature profile deviating
significantly from the monotonic profile of a vacuum black hole.  Once
the shell is accreted, the solution is once again dominated by the
curvature of the final vacuum black hole spacetime.  }}
\label{fig:sga2}
\end{figure}

\begin{figure}
\centerline{\psfig{figure=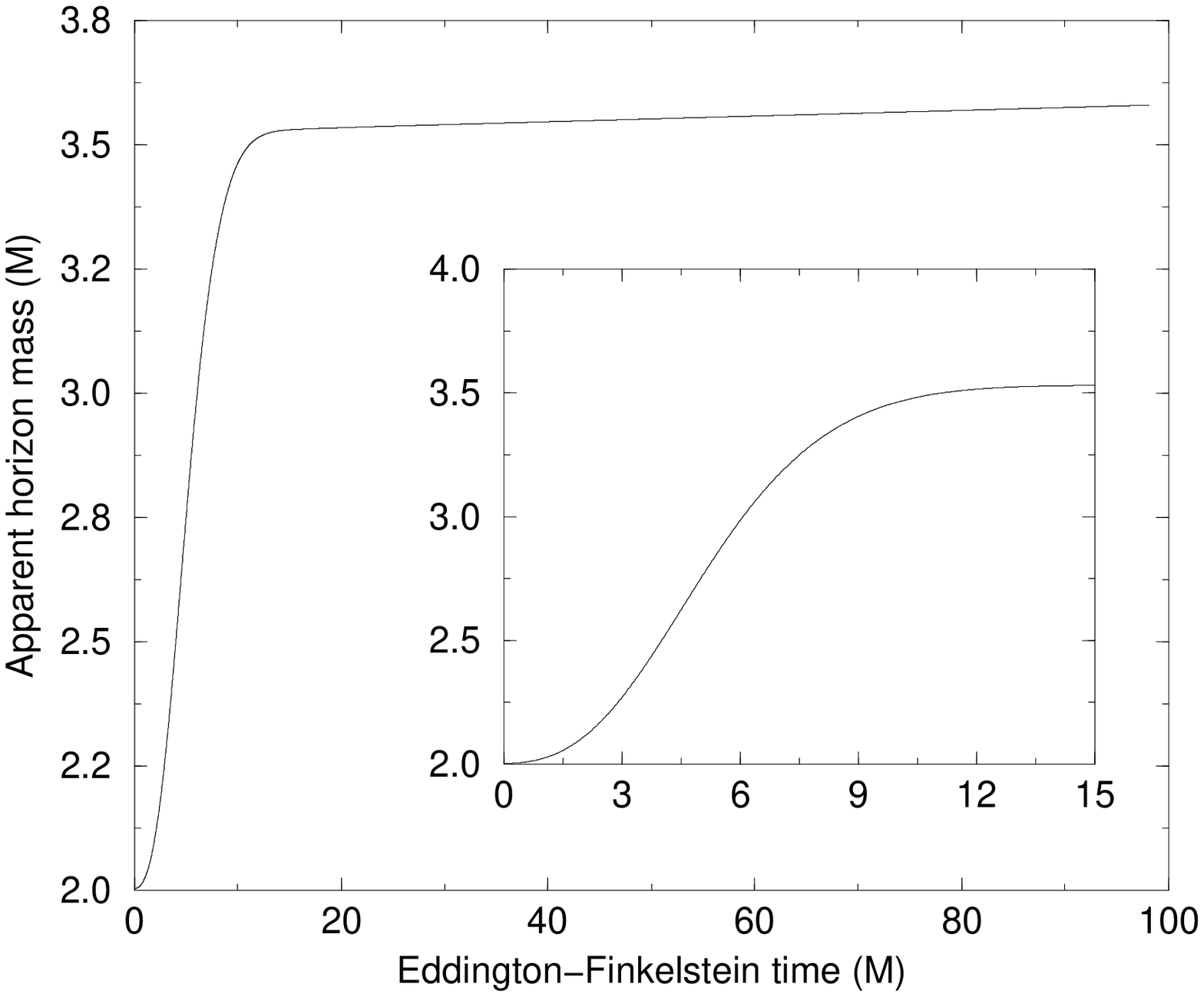,width=5.0in,height=3.8in}}
\caption{{ Spherical accretion of a self-gravitating perfect fluid:
evolution of the black hole apparent horizon mass. The mass of the apparent
horizon shows a rapid increase in the first $15M$ (enlarged in the
insert), in coincidence with the most dynamical accretion phase. The
slow, quasi-steady growth at later times is the quiet response of the
black hole to the low mass accretion rate imposed at the world tube.
This rate can be made negligibly small (if desired). }}
\label{fig:sga3}
\end{figure}

The results of the simulation are plotted in Figs.~\ref{fig:sga1},
\ref{fig:sga2} and~\ref{fig:sga3}. In Fig.~\ref{fig:sga1} we display
the evolution of the primitive variables $(\rho,u^{r},\epsilon)$, from
top to bottom, as a function of the ingoing Eddington-Finkelstein
radial coordinate, $r$.  The configuration is radially advected
(accreted) towards the hole in the first $10M-15M$.  Once the bulk of
the accretion process ends, we are left with a quasi-stationary
background solution (basically equivalent to the Bondi solution).  In
Fig.~\ref{fig:sga2} we plot the evolution of the logarithm of the
Riemann scalar curvature. We note that the initial shell has
associated with it a non-negligible curvature. At late times the
solution is again dominated by the curvature of the central black
hole.

The location of the apparent horizon of the black hole can be easily
computed during the evolution. For the simplified case of spherical
symmetry, this location is just given by the zero of the $g_{00}$
metric component. Our results show that the accretion process
initiates a rapid increase of the mass of the apparent horizon.  This
is depicted in Fig.~\ref{fig:sga3}. The horizon almost doubles its
size during the first $10M-15M$ (this is enlarged in the insert of
Fig.~\ref{fig:sga3}).  Once the main accretion process has finished,
the mass of the horizon slowly increases, in a quasi-steady manner,
whose rate depends on the mass accretion rate imposed at the
world-tube, $\Gamma$, of the integration domain.  The numerical
solution can be evolved as far as desired into the future. 

\section{Summary and Conclusions}
\label{sec:summary}

The equations of general relativistic hydrodynamics have been written
directly in terms of components of the fluid fields in an arbitrary
coordinate patch. The system of equations has been diagonalized in the
general case, aiming at the subsequent use of advanced numerical
methodology. The conservation form of the equations is preserved to
the degree possible, whereas the formulation does not require a
spacelike foliation for its implementation.

We may ask whether there is any unexpected element in the developments
here. The formulation of the hydrodynamic equations as a Cauchy
problem, (well) posed on a null spacetime surface, is intuitively
expected and has been formulated before as mentioned in the text. The
writing of the equations as a system of conservation laws on an
arbitrary foliation is implicit in their abstract representation. It
is further expected that for any non-degenerate choice of primitive
fields, the system will be hyperbolic and diagonalizable. What does
appear to be more of a happy algebraic coincidence rather than a
general feature is the possibility of {\em explicitly} solving for the
eigenfields.  Indeed, choices of variables close to the ones presented
here do not allow explicit diagonalization. We are aware of two cases
in the literature where similar explicit resolutions as the ones
reported here have been achieved. In the first case the authors made
an explicit assumption of a spacelike foliation~\cite{betal97} (see
also~\cite{inrg2}), while the second~\cite{em95} appears motivated by
a choice of variables specific to a non-relativistic Riemann solver
(the Roe solver), and leads to expressions considerably more
complicated than the ones presented here.

We have developed a number of test-beds by recomputing known solutions
in our coordinates. The performance of the algorithm was satisfactory
in each instance, which establishes the overall feasibility of the
approach. The practical value of this demonstration is that advanced
schemes developed for the hydrodynamic equations by a large community
of researchers (encompassing computational astrophysics) will be
available with minimal modifications to this more specialized field. A
further point of interest is that it is demonstrated here, implicitly,
that the Newtonian pedigree of the field of hydrodynamics is
essentially bypassed in the formulation of the relativistic equations
as conservation laws.  Our variables have no connection, in the null
case, to the instantaneous rest frame (Eulerian) observers defined by
the normal to a spacelike hypersurface.

The selection of computations in the last section is geared towards
highlighting the suitability of our approach to the study of black
holes interacting with matter. In a representative computation, we
have shown how a non-trivial increase of the mass of the black hole
horizon can be achieved naturally in the present framework. This can
be contrasted with the considerably harder task of achieving long term
black hole evolutions in a spacelike approach, as it is seen, for
example, in~\cite{magor}.

Extending the present setup to three-dimensional spacetimes appears to
be an important target for the near future. In this respect, the
feasibility of extending the vacuum CIVP with the inclusion of matter
sources has recently been reported~\cite{bishop99}.

\section{Acknowledgments}

We thank Ed Seidel for supporting this project since its inception and
Jos\'e M$^{\underline{\mbox{a}}}$. Ib\'a\~nez for many useful
discussions. Most computations were performed at the Origin 2000
supercomputer of the AEI. P.P would like to thank SISSA for
hospitality while parts of this work were being completed. J.A.F
acknowledges financial support from a TMR grant from the European
Union (contract nr. ERBFMBICT971902).


\newpage 

\newpage

\newpage

\newpage

\newpage

\newpage

\newpage

\newpage

\newpage

\begin{references}
\bibitem{projection}
Contrast those definitions with the usual spacelike approach, in which
the normal to a time-slice along with the projection operator are
applied to the stress-energy tensor to obtain its Eulerian,
``instantaneous rest frame'', components.
\bibitem{non-conservation}
We make here the clarification that the name conservation law, used
for the system of equations~(\ref{eq:cons-law}) is a well established
misnomer, as the quantities ${\bf U}^{B}$ are not strictly conserved
in a general spacetime.
\end{references}
\end{document}